\documentclass[aps, prd, 10pt, twocolumn, superscriptaddress,noshowpacs, preprintnumbers, longbibliography,nofootinbib,bibnotes,floatfix]{revtex4-1}
\usepackage[T1]{fontenc}

\usepackage{amsmath}
\usepackage[percent]{overpic}
\usepackage{caption}
\usepackage[caption=false]{subfig}
\usepackage{amsfonts}
\usepackage{comment}
\usepackage{amssymb}
\usepackage{booktabs}
\usepackage{bbold}
\usepackage[percent]{overpic}
\usepackage{epsfig}
\usepackage{graphicx}
\usepackage[caption=false]{subfig} 
\usepackage{bm}
\usepackage{array}
\usepackage{listings}
\usepackage{color}
\usepackage{url}
\usepackage{breqn}
\usepackage{makecell}
\usepackage{ulem}

\usepackage{lineno}
\usepackage{lipsum}
\usepackage{physics}
\usepackage{mathtools}

\usepackage{tabularx}
\usepackage[export]{adjustbox}
\usepackage{multirow}
\usepackage{soul}
\usepackage{xcolor}
\usepackage[utf8]{inputenc}
\usepackage{accents}
\usepackage{tikz}

\definecolor{Wildstrawberry}{rgb}{1.0, 0.26, 0.64}

\definecolor{lime}{HTML}{A6CE39}
\DeclareRobustCommand{\orcidicon}{\hspace{-1mm}
	\begin{tikzpicture}
	\draw[lime, fill=lime] (0,0) 
	circle [radius=0.16] 
	node[white] {{\fontfamily{qag}\selectfont \tiny \,ID}};
	\draw[white, fill=white] (-0.0525,0.095) 
	circle [radius=0.007];
	\end{tikzpicture}
	\hspace{-3mm}
}

\foreach \x in {A, ..., Z}{\expandafter\xdef\csname orcid\x\endcsname{\noexpand\href{https://orcid.org/\csname orcidauthor\x\endcsname}
			{\noexpand\orcidicon}}
}

\newcommand{\affiliationA}{Instituto de F\'isica Gleb Wataghin, Universidade Estadual de Campinas, R. S\'ergio Buarque de Holanda, 777, Campinas, São Paulo, Brazil} %
\newcommand{\affiliationB}{Centro de Ciências Naturais e Humanas, Universidade Federal do ABC, Avenida dos Estados, 5001, Santo André, 09280-560, São Paulo, Brazil} 
\newcommand{\affiliationC}{Universidade Estadual de Londrina, Londrina, Brazil} %

\usepackage{hyperref}

\hypersetup{
    colorlinks=true,
    linkcolor=blue,
    urlcolor=blue,
    citecolor=red,
    bookmarks=true,
    bookmarksnumbered=true,
    breaklinks=true,
    pdfstartview=FitBH
}

\begin{document}

\title{Disentangling Charged-Current Scalar NSI from the Solar Sector at JUNO}

\author{Adriano Cherchiglia \orcidA{}}
\affiliation{\affiliationA}
\affiliation{\affiliationB}
\author{Pietro Chimenti\orcidB{}}
\affiliation{\affiliationC}
\author{Christiane Frigerio Martins
\orcidE{}}
\affiliation{\affiliationC}
\author{Guilherme A. Nogueira\orcidC{}}
\affiliation{\affiliationA}
\author{Pedro Pasquini\orcidD{}}
\affiliation{\affiliationA}
\author{Orlando L.G. Peres\orcidF{}}
\affiliation{\affiliationA}


\date{\today}

\begin{abstract}

We constrain charged-current scalar non-standard interaction  with reactor antineutrinos within a quantum field theory framework of neutrino oscillations. We derive an analytical expression up to $\mathcal{O}(\varepsilon^{4})$ for the electron antineutrino survival probability including scalar NSI contributions, and show that these effects induce spectral distortions that admit a large mixing angle solution for $\theta_{12}$. Using the 59.1-day and preliminary 207.2-day JUNO datasets, we obtain the first neutrino-oscillation determination of NSI parameters individually, rather than in a flavor-summed combination, from reactor experiments. Assuming one year of data taken, we show that the only solar sector analysis  breaks the degeneracy with the solar mixing angle at $3\sigma$ of C.L.

\end{abstract}

\maketitle
\section{Introduction}
\label{sec:introduction}

Reactor antineutrinos have played a central 
role in the understanding of neutrino 
oscillations through the measurement of the 
solar mixing parameters, $\theta_{12}$ and 
$\Delta m^2_{21}$ \cite{KamLAND:2004mhv} and 
the  mixing angle, $\theta_{13}$, 
\cite{DayaBay:2012fng, Matsubara:2012uh}. 
With improved detection technology, capable of 
reaching energy resolution of $\sim 
3\%/\sqrt{E_{\text{vis}}(\mathrm{MeV})}$ and a 
medium-baseline of $\sim 52.5$~km, the Jiangmen 
Underground Neutrino Observatory (JUNO)  
reactor antineutrino experiment is 
designed to achieve unprecedented precision on 
the solar and the atmospheric oscillation 
parameters~\cite{JUNO:2015zny}. The experiment 
detects anti-neutrino events from multiple 
reactors and is sensitive to the interference 
between oscillations driven by the solar and 
atmospheric mass-squared splittings acting on 
the electron antineutrino disappearance 
channel. Such unique setup enables for a sub-percent precision measurement on these standard 
oscillation parameters. In fact, the first 59.1-day JUNO dataset already confirms the 
unprecedented sensitivity obtaining world 
leading result for $\sin^2 \theta_{12} = 0.3092 
\pm 0.0087$ and $\Delta m_{21}^2 = (7.5\pm 
0.12)\times 10^{-5}$ eV$^2$ \cite{JUNO:2025gmd}. Moreover, the preliminary result 
of 207.2-days improves the precision to $\sin^2 
\theta_{12} = 0.3036 \pm 0.0064$ and $\Delta 
m_{21}^2 = (7.388\pm 0.078)\times 10^{-5}$ 
eV$^2$ and is starting to measure the $\Delta 
m_{31}^2$ mass splitting 
\cite{JUNONEUTRINO2026}.

The JUNO experiment is also capable of 
probing physics beyond the standard 
oscillation paradigm. For the full 6 years run 
of the experiment, the high statistics will 
provide tests for unitarity violation of the 
neutrino mixing matrix \cite{Huang:2025znh}, 
neutrino magnetic moment 
\cite{Ventura:2025gfy}, quantum 
decoherence \cite{DeRomeri:2023dht}, and
renormalization group running of neutrino 
mixing parameters \cite{Ge:2024ibn,
Ge:2026hsh,Denton:2026mkp}. Interestingly, 
the first few days of data release already 
allows for testing sterile neutrinos 
\cite{Flores:2026vbx}, damping effects 
\cite{Beccaria:2026ous}, Lorentz Invariance 
violation \cite{Araya-Santander:2025jfd}, and 
time dependent neutrino masses 
\cite{Alves:2026ydc}.

Another  possibility is the presence 
of charged-current non-standard interaction (CC–NSI)~\cite{Gonzalez-Alonso:2026sgl}. In neutrino 
oscillation experiments, the effect of NSI 
 is consistently captured by a quantum field theory (QFT) 
description, in which production, propagation, 
and detection are treated as a unified process 
at the amplitude level 
\cite{Falkowski:2019kfn}. This  
approach extends the weak left-handed 4-fermion 
interaction by including non-standard 
left-handed ($L$), right-handed ($R$), scalar ($S$), 
pseudo-scalar ($P$), and tensor ($T$) 
currents. The associated 
normalized Wilson coefficients, 
$\varepsilon^X_{\alpha\beta}$, parametrize 
possible deviations from the SM interaction 
while allowing flavor-violating charged-current 
interactions. Interestingly, each new 
current contributes in a different manner for a particular process. For example, the parity of the pseudoscalar current allows $\varepsilon^P_{\alpha\beta}$ to contribute to pion decays, e.g., $\pi^+ \rightarrow \ell_\alpha + \nu_\beta$. Although it also induces a contribution to reactor $\beta$ decays, this effect is suppressed by the factor $g_P m_e/m_p \sim 0.1$ and, consequently, pseudoscalar interactions are neglected throughout this work~\cite{Falkowski:2019xoe}. Meanwhile the
left-handed NSI simply modifies the standard interaction as $\delta_{\alpha \beta} \rightarrow \delta_{\alpha \beta} + 
\varepsilon^L_{\alpha\beta}$. 
In summary, vector and axial-vector ($L/R$) NSI effects share the SM Lorentz structure and are largely absorbed into effective redefinitions of $g_V, g_A$, whereas scalar and tensor charged-current interactions remain physically distinct~\cite{Gonzalez-Alonso:2018omy, Cirigliano:2013xha}. These interactions bring an explicit  $m_e/E_e$ dependence arising from the lepton-chirality flip, as highlighted in Ref.~\cite{Falkowski:2019xoe}. Since $E_e$ lies in the few-MeV range at reactor experiments, this term grows sharply toward low energies, a regime made accessible by the energy resolution of JUNO experiment~\cite{JUNO:2025fpc}.

The scope of this work relies on the charged-current scalar non-standard interaction (CC-SNSI), such as within this scenario the oscillation probability is  modified by factor up to quartic order in $|
\varepsilon^S_{\alpha\beta}|$ that are often 
ignored in the literature. Since the bounds to 
the scalar parameters are of $\mathcal O(1)$, 
the quadratic term still 
plays a important role. Specially, we show 
that the presence of $\varepsilon^S_{e\mu}$ 
or $\varepsilon^S_{e\tau}$ significantly 
enlarge the allowed range of $\sin^2 
\theta_{12}$. Also, we show that a 10 years run 
of the JUNO experiment is capable of improving 
the current bounds to $|\varepsilon^S_{e\mu}|,
|\varepsilon^S_{e\tau}|\lesssim 5\times 
10^{-2}$ effectively excluding the large values 
of $\sin^2 \theta_{12}$ and reaching the linear regime approximation in $\varepsilon^S_{\alpha \beta}$.

This paper is organized as follows. In Sec.~\ref{sec:formalism} we present
the theoretical framework, deriving the QFT-based transition rate with SNSI at
production and detection. In Sec.~\ref{sec:simulation} we describe the JUNO
simulation, the treatment of systematic uncertainties, and the statistical
analysis.  In Sec.~\ref{sec:discuss} we present our numerical results: the constraints
on $\varepsilon^S_{e\mu}$ and $\varepsilon^S_{e\tau}$ from the
59.1-day and preliminary 207.2-day JUNO datasets, compared with existing
reactor bounds, together with a projection for a JUNO-like experiment
restricted to the solar sector, which yields a conservative estimate of the constraints on the scalar Wilson coefficients. We
conclude in Sec.~\ref{sec:conclusion}. Appendix~\ref{app:cc-nsi} contains the generalization to other NSI interactions and matter effects.

\section{Framework}
\label{sec:formalism}

At characteristic energies for reactor neutrinos, the Lee--Yang effective Lagrangian
\cite{Lee:1956qn, DayaBay:2024hya} provides an
appropriate description of these new interactions in terms of the
nucleon $(p,n)$ and fermion $(\ell_{\alpha}, \nu_\beta)$ fields.  We focus on the scalar NSI Lagrangian as given by
\begin{equation}
\mathcal{L}_{\mathrm{NSI}} \supset -\sqrt{2} G_{F}V_{ud} g_{S}\varepsilon^{S}_{\alpha\beta} (\bar{p} n)(\bar{\ell}_{\alpha} P_{L}\nu_{\beta}) + \text{h.c.},
\label{eq:LYlagrangian}
\end{equation}
where $V_{ud}$ is the relevant CKM matrix 
element and $G_{F}$ is the Fermi constant. The 
coefficient $g_{S}$ is the scalar nucleon form 
factor encoding non-perturbative QCD effects~\cite{Gonzalez-Alonso:2018omy} 
and $\varepsilon^S_{\alpha\beta}$ parametrizes 
the size of the CC-SNSI interaction in terms of the standard weak 
interaction.

In the QFT approach for neutrino oscillation, the rate of detected events for the process $\nu_\alpha \rightarrow \nu_\beta$ is given by \cite{Falkowski:2019kfn},
\begin{equation}
R_{\alpha\beta} = \frac{\sum_{k,l} e^{-2i \Delta_{kl}} \, \mathcal{P}_{\alpha}^{kl} \, \mathcal{D}_{\beta}^{kl}}{\sum_k \mathcal{P}_{\alpha}^{kk} \sum_l \mathcal{D}_{\beta}^{ll}},
\label{eq:Pab_QFT}
\end{equation}
where $\Delta_{kl}\equiv\Delta m^2_{kl}L/4E_\nu$ is the oscillation phase
 with $E_\nu$ denoting the neutrino energy, $L$ the distance between production ($\mathcal{P}$) and detection ($\mathcal{D}$), and the mass-squared difference $\Delta m^2_{kl} \equiv m_k^2 - m_l^2$. The production and detection matrix squared factors are defined, respectively, as
\begin{equation}
\hspace{-0.2cm}
\mathcal{P}_{\alpha}^{kl} = 
\int d\Pi_{\mathcal{P}} \, \mathcal{M}^{\mathcal{P}}_{\alpha k} \mathcal{M}^{\mathcal{P}*}_{\alpha l},
\text \quad 
\mathcal{D}_{\beta}^{kl} = \int d\Pi_{\mathcal{D}} \, \mathcal{M}^{\mathcal{D}}_{\beta k} \mathcal{M}^{\mathcal{D}*}_{\beta l}.
\end{equation}
with $d\Pi_{\mathcal{P/D}}$ the corresponding phase-space integration elements.

 Production of antineutrinos is taken to proceed predominantly through Gamow-Teller transitions, while both Fermi and Gamow-Teller contributions are retained at detection via IBD \cite{Chaves:2021kxe}. The presence of NSI in Eq.(\ref{eq:LYlagrangian}) modifies the standard survival antineutrino probability, $ P(\bar{\nu}_{e} \rightarrow \bar{\nu}_{e})$, into a pseudo-probability by introducing extra matrix elements \cite{Cherchiglia:2023ojf,Cherchiglia:2025dnd}. 
For example, the production matrix element is modified from the usual left-handed one $\mathcal{M}^{\mathcal{P}}_{\alpha k} = U_{\alpha k}^*\mathcal{M}^{\mathcal{P}}_L$ to a sum of left-handed and scalar contributions, $\mathcal{M}^{\mathcal{P}}_{\alpha k} = U_{\alpha k} \mathcal{M}^{\mathcal{P}}_L + [\varepsilon^S\cdot U]^*_{\alpha k} \mathcal{M}^{\mathcal{P}}_S$, and similarly for detection. Then, the change in the oscillation probability is modulated not only by the $\varepsilon^S_{\alpha \beta}$ parameters, but in principle also by  production ($p_{SS}, p_{SL}$) and detection $(d_{SL}, d_{SS})$ coefficients defined as \cite{Falkowski:2019xoe, Falkowski:2019kfn},
\begin{equation}
p_{SY} \equiv \frac{\int d\Pi_{\mathcal{P}}\mathcal{M}^{\mathcal{P}}_{S}\, \mathcal{M}^{\mathcal{P}\,*}_{Y}}{\int d\Pi_{\mathcal{P}}\lvert \mathcal{M}^{\mathcal{P}}_{L} \rvert^{2}}, \qquad d_{SY} \equiv \frac{\int  d\Pi_{\mathcal{D}} \mathcal{M}^{\mathcal{D}}_{S}\, \mathcal{M}^{\mathcal{D}\,*}_{Y}}{\int  d\Pi_{\mathcal{D}} \lvert \mathcal{M}^{\mathcal{D}}_{L} \rvert^{2}},
\end{equation}
with $Y = S, L$. These amplitude ratios between NSI and SM contributions in production and detection were evaluated for the scalar and other interactions in Ref.~\cite{Falkowski:2019xoe}. The absolute value of these coefficients exhibits an energy dependence and is of order $\sim 0.3$ near the IBD threshold, as shown in Ref.~\cite{Gonzalez-Alonso:2026sgl}. With the detection and production coefficients we obtain the pseudo-oscillation probability in terms of the usual  antineutrino transition amplitude $\bar{S}_{\alpha \beta} = \sum_{k} e^{-i (m_{k}^{2}/2 E) L} \, U_{\alpha k} U_{\beta k}^{*}$, with $U$ the PMNS matrix, 

\begin{widetext}
\begin{align}
\frac{R_{ee}}{\phi_e^{\text{SM}} \sigma_e^{\text{SM}}} &= P(\bar{\nu}_{e} \rightarrow \bar{\nu}_{e}) + 2 d_{\rm SL} \,\Re\!\left( \varepsilon_{e\ell} \bar{S}_{ee}^* \bar{S}_{e\ell} \right) + (d_{\rm SS} +  \, p_{\rm SS}) |\varepsilon_{e\ell}|^2 \, P(\bar{\nu}_{\ell}\rightarrow \bar{\nu}_{e}) \nonumber\\
&\quad  + 2  \, d_{\rm SL} \, p_{\rm SS} |\varepsilon_{e\ell}|^2 \Re\!\left( \varepsilon_{e\ell} \bar{S}_{e \ell}^* \bar{S}_{\ell\ell} \right)  + \,d_{\rm SS}\, p_{\rm SS} |\varepsilon_{e\ell}|^4 \, P(\bar{\nu}_{\ell} \rightarrow \bar{\nu}_{\ell}) \, .
\label{eq:PeeNSI}
\end{align}
\end{widetext}
Here we omit the scalar superscript, and $\varepsilon_{e\ell}$ is understood to denote $\varepsilon_{e\ell}^{S}$ throughout, since only scalar interactions are considered in what follows.An explicit derivation of Eq.~(\ref{eq:PeeNSI}) and its generalization to other NSIs  are given in Appendix~\ref{app:cc-nsi}.

Within the adopted framework a few remarks are in order:
\begin{itemize}
    \item We adopt the hypothesis that production is dominated by the Gamow--Teller interaction, so that $p_{SS}\approx 0$. Indeed, roughly $70\%$ of the beta-decay strength proceeds via Gamow--Teller transitions, with the remainder governed by first-forbidden transitions, whose theoretical uncertainties are large~\cite{Falkowski:2019xoe}. Under this assumption the second line  of Eq.~\eqref{eq:PeeNSI} vanishes, and the pseudo-probability reduces to $\mathcal{O}(\varepsilon_{e \ell}^{2})$.
 \item The recent NSI constraints from reactor experiments \cite{Falkowski:2019xoe,Chaves:2021kxe} places the CP violation in the $2$--$3$ sector rather than in the conventional $1$--$3$ sector, absorbing the $\theta_{23}$ and CP-phase dependence into effective coefficients $\tilde{\varepsilon} = \sum_{\ell} g_{\ell}\, \varepsilon_{e \ell}$, with flavour-dependent weights $g_{\ell}\equiv g_{\ell}(\theta_{23},\delta_{CP})$.  We constrain $\varepsilon_{e\mu}$ and $\varepsilon_{e\tau}$ individually, setting all other coefficients of Eq.~(\ref{eq:LYlagrangian}) to zero. Since
    the 59-day JUNO dataset is not  sensitive to the atmospheric-driven oscillation pattern, so our estimates remain confined to the solar sector. 
 \item The scalar contributions from $\varepsilon_{e\mu}$ and $\varepsilon_{e\tau}$ enter incoherently: were both flavors switched on simultaneously, they would add as a sum of probabilities rather than amplitudes. Throughout this work we restrict our analysis to one lepton flavor $\ell$ at a time. Since this new physics is encoded entirely at the production and detection vertices, leaving matter interactions unchanged, a consistent treatment of matter effects could be incorporated in our framework perturbatively via the matter potential, following the semi-analytical approach of Ref.~\cite{Khan:2019doq}.
\end{itemize}

\section{Simulation of JUNO experiment}
\label{sec:simulation}

We use the \texttt{GLoBES}~\cite{Huber:2004ka, 
Huber:2007ji} library to simulate the JUNO 
experiment and neutrino propagation by substituting 
the standard oscillation probability by the one 
in Eq.~(\ref{eq:PeeNSI}). We modeled the effect 
of the detector by including our own reconstruction 
matrix $R_{ij}$ into GLoBES, where $j$ labels the true 
neutrino energy and $i$ the reconstructed prompt
energy bin. Each column of $R_{ij}$ represents the 
reconstructed energy distribution for events generated 
in one true-energy bin. With this definition, the 
predicted reconstructed spectrum is as follows
\begin{equation}
   N_i^{\mathrm{rec}}
 =
   \sum_{j={\rm true~bins}}
   R_{ij}
   \,N_j^{\mathrm{true}}\,.
\end{equation}
We calculated the column histogram of $R_{ij}$
via a Monte Carlo simulation with the event rate $10^9$ for each true energy $E_j$ such, that
\begin{align}
    R_{ij}
= 
  \frac{\# \nu~with~E_j~reconstructed~with~E_i^{\rm rec}}{\# \nu~with~E_j}.
\end{align}
The reconstructed visible energy $E_i^{\rm rec}$ is 
obtained by several reconstruction steps. First we 
randomly sampled positron energies from the interval 
$E_{e^+} \in [E_{\rm min}, E_{\rm max}]$, where 
$E_{\rm min}, E_{\rm max} \sim E_\nu - 0.784$ MeV, are 
the minimum and maximum allowed values as given in 
Eq.(12) of \cite{Strumia:2003zx}. Then, a true 
prompt energy is defined as $E_{\rm pr} = E_{e^+} + 
m_e$, with $m_e$ the electron mass, corresponding to 
the electron-positron annihilation at the detector. 
Next, we randomize a reconstructed prompt energy 
$E_{\rm pr}^{\rm rec} $ via a Gaussian distribution 
with central value $E_{\rm pr}$ and energy resolution 
parametrized according to $\sigma/E_{pr} = 
\sqrt{a^2/E_{\rm pr} + b^2}$ (with $E_{\rm pr}$ in MeV 
and $a = 3.3\%$ and $b = 1\%$) 
as given in \cite{JUNO:2025fpc}. We also take into 
account the non-linearity of the detector by 
defining our final reconstructed energy $E^{\rm rec} 
\equiv f_{\rm NL}(E_{\rm pr}^{\rm rec}) E_{\rm pr}^{\rm 
rec}$ where the function $f_{\rm NL}$ was 
extracted from Figure(6d) of \cite{JUNO:2025fpc}. We 
verified that the analytical approach of 
Ref.~\cite{Capozzi:2013psa} to obtain $R_{ij}$ 
yields a similar result compared to our 
simulations.
\begin{align}
  N_j^{\mathrm{true}}
=
  \sum_{r = {\rm reactors}} \sigma(E_j) P_{\bar \nu_e \rightarrow \bar \nu_e}(L^r) \phi^r_{\bar \nu_e}(L^r)
\end{align}
\begin{figure*}[t!]
    \centering
    \includegraphics[width=0.49\linewidth]{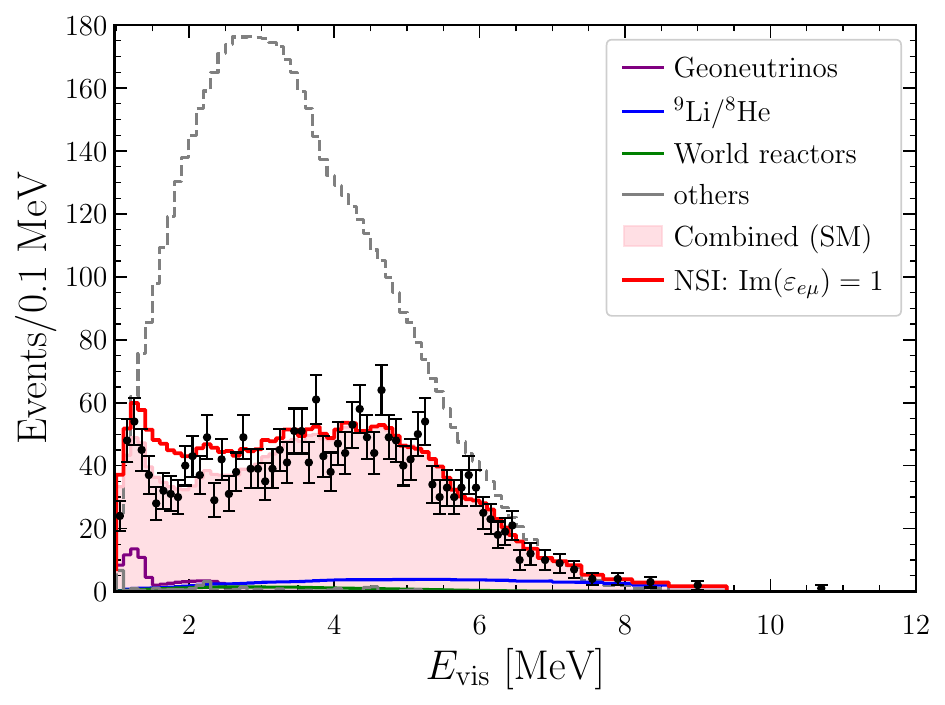}
     \includegraphics[width=0.475\linewidth]
     {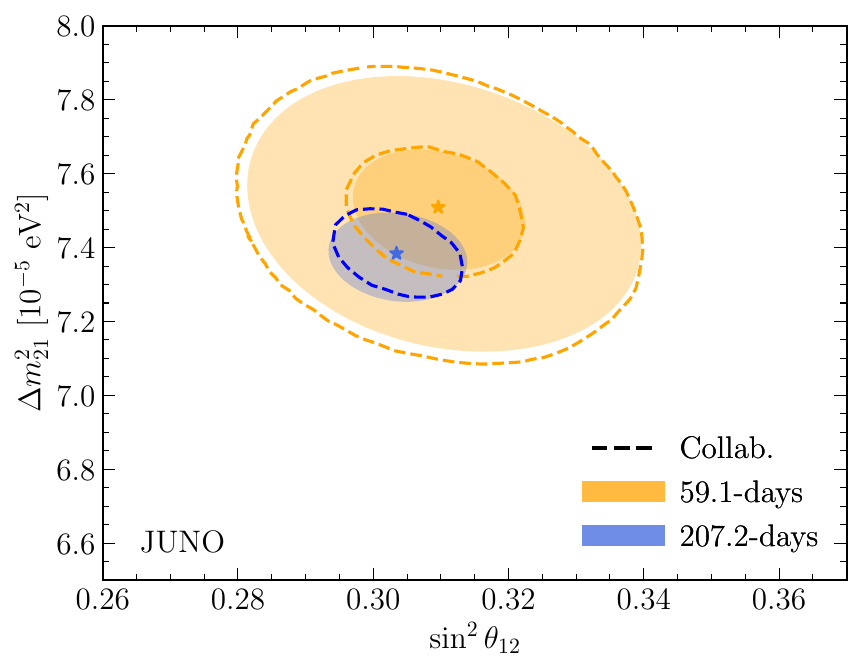}
  \caption{{\bf Left:} Predicted event spectra (solid curves) compared with the
JUNO collaboration result (black dashed histogram), showing excellent
agreement. The red curve includes CC-SNSI with
$\mathrm{Im}\,\varepsilon^S_{e\mu} = 1$ and $\mathrm{Re}\,\varepsilon^S_{e\mu} = 0$, while the shaded region
corresponds to the standard oscillation prediction with all backgrounds combined.
{\bf Right:} Comparison between sensitivity of JUNO(dashed lines) and our simulation (filled regions) for the 59.1-days analysis (orange) and the 207.2-days analysis. The regions are for $1\sigma$ and $3\sigma$, however, there is no 3$\sigma$ region for the 207.2 analysis in the collaboration result, so we omit it in our region for easy readability of the image.
}
\label{fig:event_rates}
\end{figure*}
The true event distribution $N_j^{\mathrm{true}}$ 
is obtained by calculating the IBD cross-section,
the flux of each reactor source, $\phi^r(L^r)$ at distance $L^r$ and the corresponding  oscillation probability $P_{\bar \nu_e \rightarrow \bar \nu_e}(L^r)$,

The flux of anti-neutrinos is a combination of 10 
reactors at various thermal powers and distances. We 
used the nominal power and locations for the 6 
Yangjiang and 2 Taishan cores ($L\approx 52.5$ km), 
the Daya-Bay reactor core ($L\approx 215$ km) as 
described in Table 1 of \cite{JUNO:2021vlw}. We also 
include the Fangchenggang power plant in 
$411.7\,\mathrm{km}$, as adopted in the updated 
version of Ref.~\cite{Esteban:2024eli}. Then, the 
total reactor flux is a sum of each individual flux 
$\phi^r$. For the model of $\phi^r(E_\nu)$ we started 
from the Huber--Mueller parametrization 
\cite{Huber:2011wv,Mueller:2011nm},
\begin{align}
    \phi^r_{\bar \nu_e}(L^r)
=
 \frac{P^r}{4\pi (L^r)^2}\sum_{k} \frac{f_k}{\bar f} {\rm exp}\left(\sum_{p=1}^6 \alpha_{pk} E^{p - 1}\right)
\end{align}
where $f_k$ ($\bar f = \sum f_k$) are the fractional 
power of each isotope, $k =$$^{235}$U,$^{238}$U, 
$^{239}$Pu, and $^{241}$Pu and $\alpha_{pk}$ are 
parameters to fit. We performed a fit by varying 
the $\alpha_{pk}$ around their best fit given in 
\cite{Mueller:2011nm} in order to better fit 
the non-oscillated event prediction from the JUNO 
collaboration result in \cite{JUNO:2025gmd}. Our 
final result gets close to the non-oscillated result,
but does not fit the well known excess at around $5$ 
MeV. So we add a pre-reconstruction bin-to-bin 
normalization $N_j^{\rm true} \rightarrow (1 + a_j) 
N_j^{\rm true} $ were $a_j$ are obtained by requiring 
a perfect match to the non-oscillated event prediction.

The main backgrounds are slightly different for each 
JUNO analysis. For the 59.1-day analysis we considered 
the backgrounds from geoneutrinos, $^9$Li/$^8$He 
decays, world reactors and grouped all other 
subdominant sources, including the 
$^{214}$Bi--$^{214}$Po background as ``others''. The 
background rates were obtained from Figure3 of 
\cite{JUNO:2025gmd}. For the 207.2 days analysis 
we considered geoneutrinos, $^9$Li/$^8$He 
decays, world reactors, $^{214}$Bi--$^{214}$Po and 
grouped all other remaining subdominant backgrounds
into ``others''. The background curves were extracted 
from the event rate figures in \cite{JUNONEUTRINO2026}.

The resulting event rates for the 59.1-day analysis 
are shown in the left panel of Figure 
\ref{fig:event_rates}, including the signal 
and background, described above, highlighted by the 
light pink curve. We also show the event rates 
considering the scalar NSI with ${\rm Im}
[\varepsilon_{e\mu}] =1 $ and ${\rm Re}
[\epsilon_{e\mu}] =0 $  in the red-solid line. The curve with scalar NSI mainly increases 
the event rate for $E_{\rm vis} < 4$ MeV. That is the
region most sensitive to variations of $\sin^2 
\theta_{12}$; consequently, the NSI can fake the 
behavior of the solar mixing angle and increase the 
allowed region for the analysis, as we show later. The right panel contrasts the standard oscillation determinations for both datasets employed, showing good agreement  with the data extracted from Ref.~\cite{JUNO:2025gmd} and Ref.~\cite{JUNONEUTRINO2026}, 
indicating promising prospects for probing NSI effects. 

We perform the sensitivity analysis in three 
scenarios: (1) The 59.1-day data set 
\cite{JUNO:2025gmd}, (2) the 207.2-day
 preliminary data set \cite{JUNONEUTRINO2026}
, and (3) a future prediction for 1 year and 10 years of data taken. 
For the 59.1 and 207.2 we always compare the 
prediction with the data points, while the 10-years 
prediction we scale the 207.2-days curves by the 
corresponding time and perform the fit by comparing 
predictions with an Asimov dataset with $\sin^2 
\theta_{12}$ (true) and $\Delta m_{21}^2$(true) 
corresponding to the best-fit of the 207.2-days 
analysis.

All sensitivities are obtained through a 
 $\chi$-squared analysis with $\chi^2$ 
function defined as
\begin{align}
  \chi^2 
= &
  \left({\bm N}^{\rm data} - {\bm N}^{\rm pred}\right)^{\!\top} V^{-1} \left({\bm N}^{\rm data} - {\bm N}^{\rm pred}\right)
\nonumber 
\\ & 
+ \left(\frac{a^2}{\sigma_a}\right)^2
+ \sum_{j=\rm bkg} \left(\frac{b^2_j}{\sigma_{b_j}}\right)^2
\label{eq:chi2_def}
\end{align}
where ${\bm N}^{\rm data} = (N^{\rm data}_1, N^{\rm 
data}_2, \dots, N^{\rm data}_n)$ is the vector 
containing the data points measured by the JUNO 
collaboration, with $n = 66$ or 156 for the 59.1-days 
and 207.2-days analysis respectively. The predicted 
events are included as ${\bm 
N}^{\rm pred} = (1+a) {\bm N}^{\rm sig} +  \sum_{bkg} 
(1 + b_{bkg}) {\bm N}^{\rm bkg}$, with $a$ 
and $b_{1}, b_{2}, \dots$ overall 
normalization nuisance parameters with errors 
$\sigma_a$ and $\sigma_{b_1}, \sigma_{b_2}, \dots$, 
respectively. The values of the errors were extracted 
from the anti-neutrino candidate summary tables for
the 52.1-days \cite{JUNO:2025gmd} and 207.2-days 
\cite{JUNONEUTRINO2026} analysis. We include the 
Daya-Bay flux measurement uncertainties 
\cite{DayaBay:2025ngb} into the 
covariance matrix $V$. The original systematic 
uncertainty from Daya-Bay is given as a matrix $S$ 
with 19 reconstructed energy bins. 
We propagate those uncertainties by constructing 
an $n\times 19$ auxiliary matrix $\overline N$ 
where each element $\overline N_{ij}$ represents
the total number of predicted signal events at 
reconstructed energy bin $i$ originated with
true energies lying within one of the 19 energy 
intervals $j$ of the original Daya-Bay analysis. 
Then, the covariance matrix in Eq.~(\ref{eq:chi2_def}) 
is given by $V_{ij} = N_i^{\rm data}\delta_{ij} + 
(\overline N \cdot S \cdot \overline N^T)_{ij} $.

The result of our fit for $\sin^2\theta_{12}$ versus 
$\Delta m_{21}^2$ can be found in the right panel of 
Figure\ref{fig:event_rates}. Notice the remarkable 
similarities of our simulation and the collaboration 
result. Specially, both the 59.1-days and the 207.2-
days analysis result in a very good approximation of 
the regions, with very little changes in the 
simulation. Specially, the main differences are the 
bin number and the background, which also changes
from both JUNO data sets.

\begin{figure*}[t]
    \centering
    \begin{overpic}[width=0.47\linewidth]{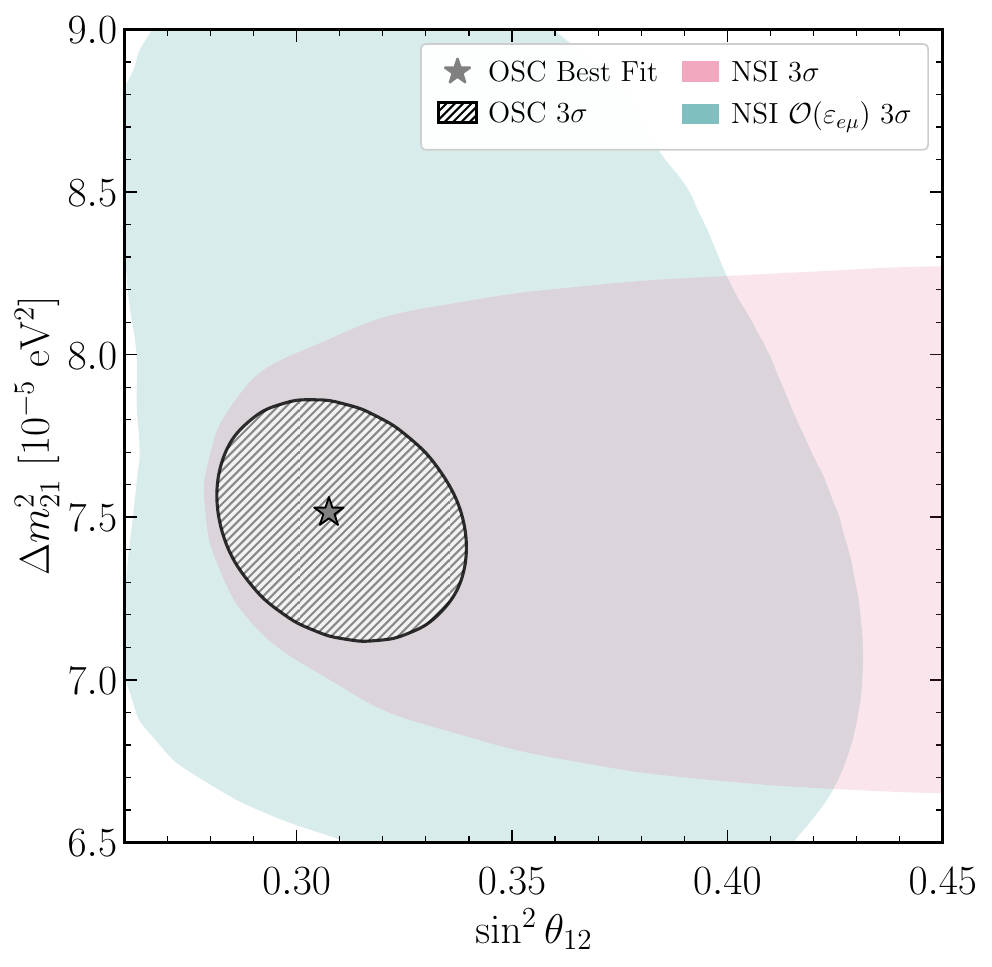}
        \put(20,85){\textbf{(a)}}
    \end{overpic}\hfill
    \begin{overpic}[width=0.47\linewidth]{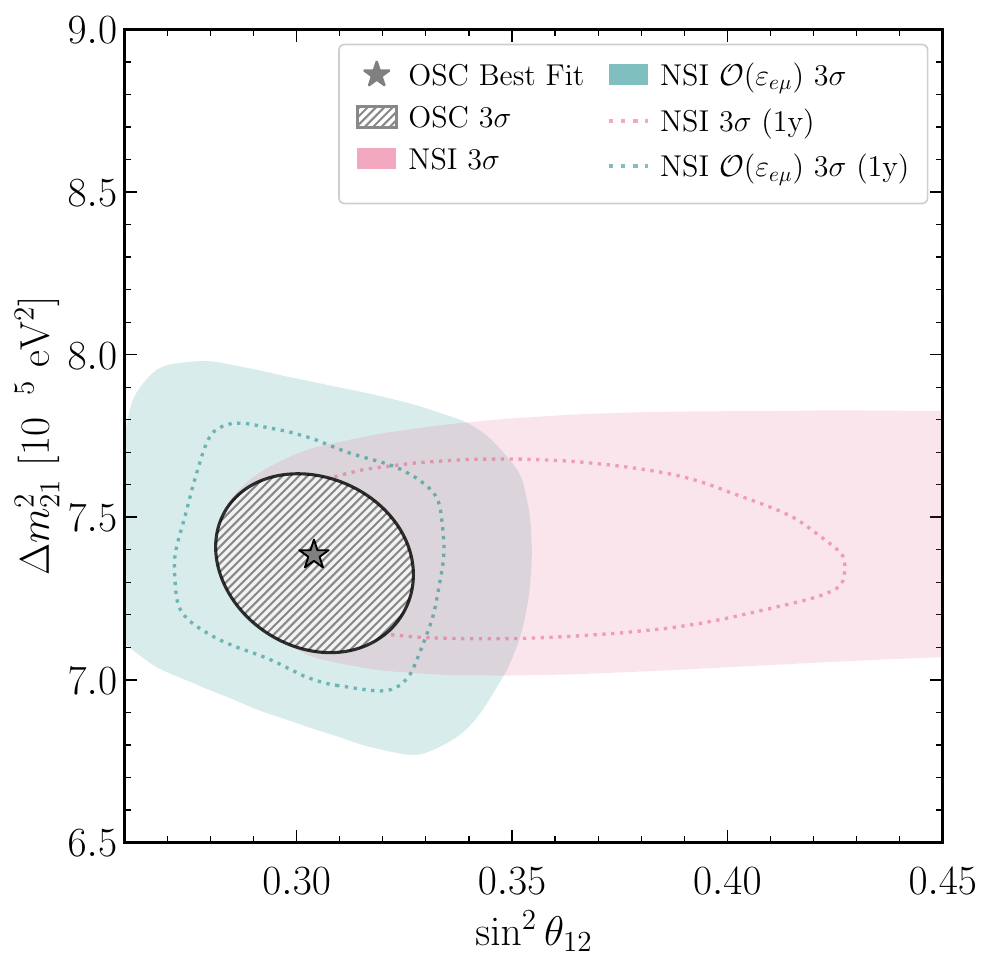}
        \put(20,85){\textbf{(b)}}
    \end{overpic}
    \caption{%
    Allowed $3\sigma$ regions in the $(\sin^{2}\theta_{12},\,\Delta m^{2}_{21})$
    plane from JUNO; stars mark the best-fit points.
    {\bf (a)}~From our $\chi^{2}$ analysis of the 59.1-day dataset: filled
    contours denote the confidence regions with NSI at linear order (teal) and at
    $\mathcal{O}(\varepsilon_{e\ell}^{2})$ (pink), while the hatched black contour (gray hatched) shows the allowed oscillation region
    {\bf (b)}~For the standard oscillation fit (gray hatched) and for fits including the
    $\varepsilon_{e\mu}$ parameter, where the pink and teal contours correspond
    to the nonlinear and linear parametrizations, respectively. Shaded regions
    use the 207.2-day dataset. Dotted contours show the projected regions for one
    year of exposure.}
    \label{fig:diffchi2}
\end{figure*}

\section{Discussion}
\label{sec:discuss}

As shown in Eq.~\eqref{eq:PeeNSI}, the scalar coefficient enters the survival probability through a term linear in $\varepsilon_{e\ell}$ and terms up to quartic order, each weighted by a different combination of the nuclear spin-dependent detection and production coefficients. Under the hypothesis of a pure Gamow--Teller contribution, one finds $p_{\mathrm{SL}}=0$ and $p_{\mathrm{SS}}\approx0$ (Sec.~\ref{sec:formalism}), so that the NSI expression reduces to $\mathcal{O}(\varepsilon_{e \ell}^{2})$. To isolate the phenomenological impact of each order on the SM solar parameters, we  examine the linear and quadratic contributions separately, taking $\ell=\mu$ as an illustrative example:
\begin{equation}
\frac{R_{ee}}{\phi_e^{\mathrm{SM}}\sigma_e^{\mathrm{SM}}} \simeq P_{ee}^{\mathrm{SM}} + N_{e\mu}^{(\varepsilon)} + N_{e\mu}^{(\varepsilon^{2})}.
\label{eq:rate_quadratic_expansion}
\end{equation}
Here, $P_{\alpha\beta}^{\mathrm{SM}}\equiv P(\bar\nu_\alpha\to\bar\nu_\beta)$ is the standard oscillation probability of Eq.~\eqref{eq:PeeNSI}, and $N_{e\mu}^{(\varepsilon_{e}^{n})}$ denotes the term of that equation with $\mathcal{O}(\varepsilon_{e \mu}^{n})$ dependence. This discrimination is useful for understanding the behaviour shown in Figure~\ref{fig:diffchi2}, which shows the allowed regions in the 
$(\sin^2\theta_{12},\,\Delta m^2_{21})$ plane derived from the 
59.1-day (a) and preliminary 207.2-day (b) datasets, together with 
the 1-year sensitivity. The 
comparison between the linear (light blue) and complete (light pink) 
expression is discussed below.

Different values of $\varepsilon_{e\ell}$ and of the solar parameters can produce similar spectral distortions, giving rise to the allowed region shown in Figure~\ref{fig:diffchi2}.  The linear term arises from the interference between the real SM amplitude and the Wilson coefficients and is therefore the origin of the phase degeneracies. This can be visualized by noting that at JUNO-like baselines the amplitude $\bar{S}_{e\mu}$ is governed by the solar phase, then writing
$\varepsilon_{e\mu}=|\varepsilon_{e\mu}|e^{i\phi_{e\mu}}$ with $\bar{S}_{ee}\simeq1$ at leading order, using the oscillation phase $\Delta_{21}$ defined in Sec.~\ref{sec:formalism} and the mixing-angle constant $\kappa\equiv c_{13}c_{12}s_{12}c_{23}$, gives
\begin{equation}
N_{e\mu}^{(\varepsilon)} \simeq 2d_{\mathrm{SL}}\,|\varepsilon_{e\mu}|\,\kappa \left[\cos(\phi_{e\mu}-\Delta_{21})-\cos\phi_{e\mu}\right].
\label{eq:linear_analytic}
\end{equation}
The leading effect of $N_{e\mu}^{(\varepsilon)}$ is thus an interference with the solar oscillation phase, making the fitted $\phi_{e\mu}$ strongly correlated with $\Delta m^{2}_{21}$, while $|\varepsilon_{e\mu}|$ is mainly degenerate with the solar mixing angle, both parameters measured with unprecedented precision by the first JUNO data~\cite{JUNO:2025gmd}. Replacing $\varepsilon_{e\mu} \to \varepsilon_{e\tau}$ drives a transition from correlated to anti-correlated behaviour in the $(\mathrm{Im}\,\varepsilon^S_{e\ell},\,\Delta m^{2}_{21})$ plane, since the two coefficients are related by the exchange $c_{23} \to -s_{23}$ in Eq.~\eqref{eq:linear_analytic}.

\begin{figure}[t!]
    \centering
    \includegraphics[width=1\linewidth]{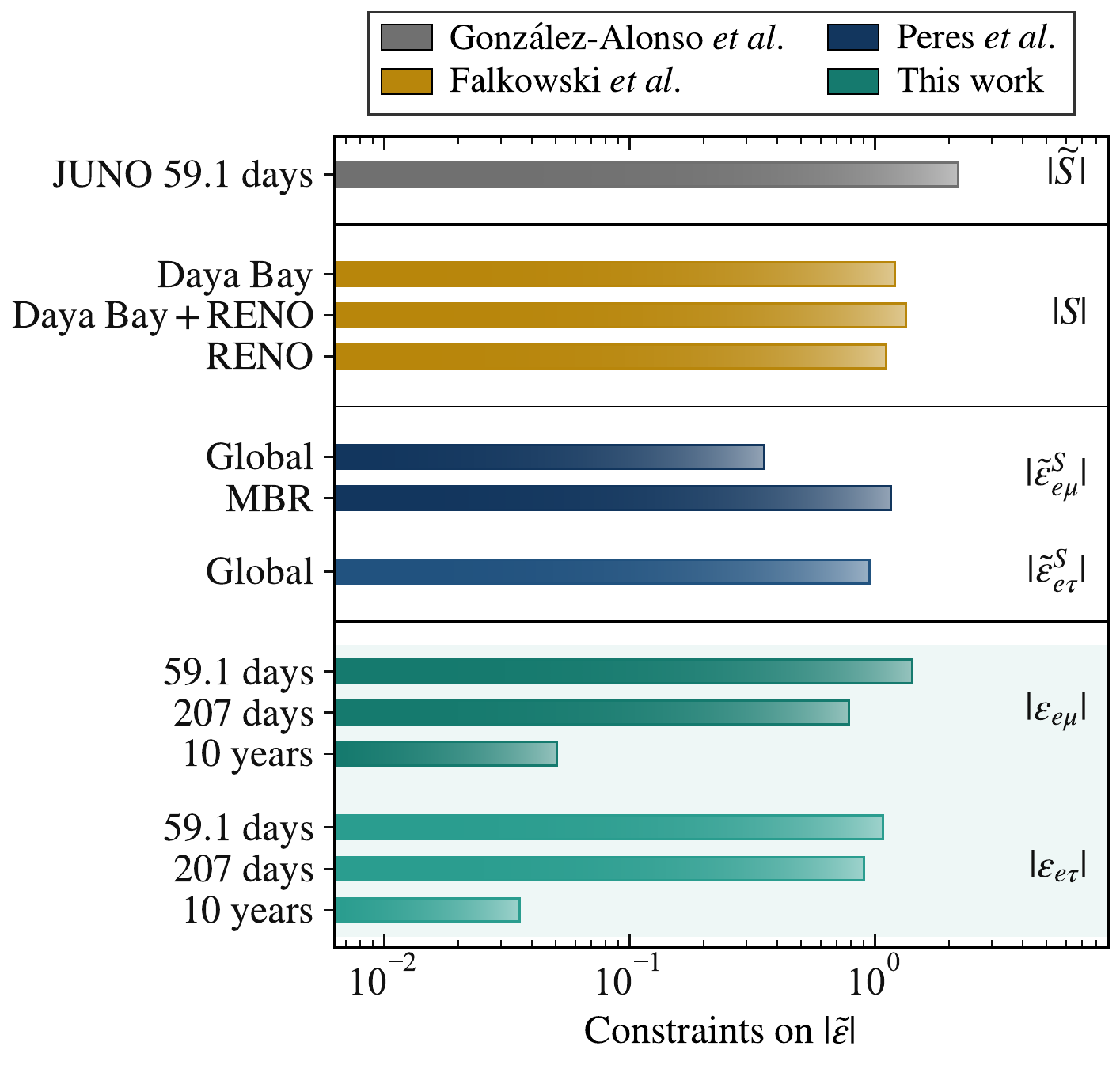}
   \caption{Upper bounds on CC-SNSI from reactor
antineutrinos at $1\sigma$. Yellow: bounds on the effective combination
$[S]$ from Daya Bay and RENO~\cite{Falkowski:2019xoe}. Blue: bounds on the
rotated couplings $\widetilde{\varepsilon}^{\,S}_{e\ell}$ from a global
analysis of $\bar{\nu}_e$ disappearance~\cite{Chaves:2021kxe}. Gray:
linear-order bound on $\widetilde{S}$ from the 59.1-day JUNO
dataset~\cite{Gonzalez-Alonso:2026sgl}. Teal: this work, constraining the
individual Wilson coefficients $\varepsilon^S_{e\mu}$ and
$\varepsilon^S_{e\tau}$ at $\mathcal{O}(\varepsilon_{e\ell}^2)$ from the 59.1-day
and 207.2-day JUNO datasets, with the 10-year projection. Bars show the $1\sigma$ bounds on $|\tilde{\varepsilon}|$, where $\tilde{\varepsilon} = \sum_{\ell} g_{\ell}\, \varepsilon_{e \ell}$, obtained from the profiled $\Delta\chi^2=1$ region in the complex plane, combining each flavor with its respective weight $g_{\ell}$. Since each analysis constrains a different combination of weights, the bars compare sensitivities to the same class of interactions rather than identical parameters.}
 \label{fig:bars}
\end{figure}

The quadratic term $N_{e\mu}^{(\varepsilon^2)}$ follows from the same solar-dominated approximation applied to $P_{\mu e}^{\mathrm{SM}}$, yielding
\begin{equation}
N_{e\mu}^{(\varepsilon^{2})} \simeq 4d_{\mathrm{SS}}\, |\varepsilon_{e\mu}|^2\, \kappa^2 \sin^2\!\left(\frac{\Delta m^2_{21} L}{4E_\nu}\right).
\label{eq:quadratic_rate}
\end{equation}
This term is independent of $\phi_{e\mu}$ and is responsible for the strong degeneracy with the mixing angle represented by the light pink region in Figure~\ref{fig:event_rates}, giving rise to the LMA solution. Although the linear term $N_{e \mu}^{(\varepsilon)}$ dominates for $\varepsilon_{e\ell}$ near zero, the 59.1- and 207.2-day JUNO datasets only constrain $|\varepsilon_{e\ell}|$ to values of $\mathcal{O}(1)$, as shown in Figure~\ref{fig:bars}. At such values the quadratic term is no longer a small correction: it is the $\mathcal{O}(\varepsilon_{e\mu}^{2})$ term that primarily shapes the allowed region, so that once the full expression is considered the degeneracy with the solar mass scale becomes weaker than Eq.~\eqref{eq:linear_analytic} alone, as a linear correction to the electron antineutrino survival probability, would suggest.

Figure~\ref{fig:diffchi2} shows that, for $\varepsilon_{e\mu}$, the sensitivity analysis restricted to the solar sector of a JUNO-like experiment already breaks the LMA-like degeneracy at $3\sigma$~C.L. with 1 year of data taken. The same mechanism operates for $\varepsilon_{e\tau}$, affecting the SM regions. We also emphasize that the analytical expressions derived thus far were obtained in the vacuum approximation in order to highlight the phenomenological behaviour of each term. The survival probability used in the fit retains the full matter-modified amplitudes $S_{\alpha\beta}$, computed numerically. 

\subsection{NSI Constraints}

The constraints  summarized in Figure~\ref{fig:bars} were obtained by scanning a four-dimensional grid in $\sin^2\theta_{12},\,\Delta m^2_{21},\,\mathrm{Re}\,\varepsilon_{e\ell}$ and $\mathrm{Im}\,\varepsilon_{e\ell}$, with the remaining oscillation parameters fixed to their global best-fit
values~\cite{Esteban:2024eli}. The $\chi^2$ is evaluated independently for the
real and imaginary parts of the Wilson coefficient, and the bounds on
$|\varepsilon_{e\ell}|$ are obtained by profiling over the solar parameters
and over the complex plane of $\varepsilon_{e\ell}$, so that the presented
limits fully account for the degeneracies discussed above.

At $1\sigma$, the 59.1-day dataset gives
\begin{equation}
   \mathrm{Re}\,\varepsilon_{e\mu} = \text{0.30}^{+1.00}_{-1.70}, \quad \mathrm{Im}\,\varepsilon_{e\mu}= \text{0.10}^{+0.90}_{-1.10},
\end{equation}
in the muon sector and
\begin{equation}
   \mathrm{Re}\,\varepsilon_{e\tau} = -\text{0.20}^{+1.20}_{-0.70}, \quad \mathrm{Im}\,\varepsilon_{e\tau}= -\text{0.10}^{+0.80}_{-0.70},
\end{equation}
in the tau sector. Adding the preliminary 207.2-day exposure tightens both
intervals by roughly a factor of two while leaving them consistent with the
59.1-day result and centered near zero, as expected in the absence of a
genuine signal:

\begin{equation}
   \mathrm{Re}\,\varepsilon_{e\mu} = \text{0.00}^{+0.68}_{-0.53}, \quad \mathrm{Im}\,\varepsilon_{e\mu} = -\text{0.08}^{+0.38}_{-0.68},
\end{equation}
in the muon sector and
\begin{equation}
   \mathrm{Re}\,\varepsilon_{e\tau} = -\text{0.30}^{+1.05}_{-0.45}, \quad \mathrm{Im}\,\varepsilon_{e\tau} = \text{0.23}^{+0.68}_{-0.53},
\end{equation}
in the tau sector. In both channels the real and imaginary parts remain
compatible with zero within $1\sigma$, and no significant deviation from the
Standard Model emerges from either exposure.

The comparison in Figure~\ref{fig:bars} must be interpreted with care, since existing bounds are quoted in different SNSI conventions. The bounds of Ref.~\cite{Falkowski:2019xoe} are likewise defined at the Lagrangian level by $\varepsilon_{\alpha\beta}$, but the quantity shown as $[S]$ is not a single Wilson coefficient: it is the effective scalar combination of $\varepsilon_{e\mu}$ and $\varepsilon_{e\tau}$ entering $\bar\nu_e$ disappearance, constrained jointly with the effective angle $\widetilde\theta_{13}$. Since $\mathrm{Re}[S]$ is partially degenerate with $\widetilde\theta_{13}$ while $\mathrm{Im}[S]$ enters through a distinct CP-odd interference term, the bounds of Ref.~\cite{Falkowski:2019xoe} constrain this combination rather than $\varepsilon_{e\mu}$ directly. The same caveat applies to Ref.~\cite{Chaves:2021kxe}, whose results are reported in terms of the rotated coupling $\widetilde\varepsilon^X = \varepsilon^X\cdot\,U(\theta_{23},\delta)$, so that $\widetilde\varepsilon^S_{e\mu}$ and $\widetilde\varepsilon^S_{e\tau}$ mix the unrotated coefficients $\varepsilon_{e\mu}$. Therefore, the Figure~\ref{fig:bars} should therefore be read as a comparison of sensitivities to the same class of  CC-SNSI, not as a direct comparison of identical parameters.

The bounds of Ref.~\cite{Gonzalez-Alonso:2026sgl}, obtained from the same
59.1-day JUNO dataset, are nominally stronger than ours. This difference is
fully accounted for by the analysis setup: Ref.~\cite{Gonzalez-Alonso:2026sgl}
works at linear order in $\varepsilon$, constrains a single rotated
combination $c_{23}\,\varepsilon_{e\mu}-s_{23}\,\varepsilon_{e\tau}$ with
one component at a time, and anchors the solar parameters with external
global-fit priors. The linear truncation, in particular, removes the
positive-definite $\mathcal{O}(\varepsilon^2)$ term responsible for the
LMA-like degeneracy, which dominates the shape of our allowed
regions.

With the adopted treatment of systematics, the currently available 59.1-day JUNO dataset already reaches a sensitivity competitive with existing reactor constraints, $|\varepsilon_{e \mu}| \le 1.41$ and $|\varepsilon_{e \tau}|\le 1.08$, tightening to $|\varepsilon_{e \mu}| \le 0.78$ and $|\varepsilon_{e \tau}|\le 0.90$ with the preliminary 207.2-day dataset. Projecting to ten years of data taking, the sensitivity improves by more than an order of magnitude relative to the 59.1-day result, $|\varepsilon_{e \mu}| \le 5.1\times10^{-2}$ and $|\varepsilon_{e \tau}|\le 3.6\times10^{-2}$, surpassing all previous determinations of CC-SNSI from reactor neutrino oscillations.

As discussed before, antineutrino production in $\beta$ decays is dominated by Gamow-Teller transitions ($\sim70\%$), so a scalar interaction generates no first-order modification of the production rate in $\varepsilon_{e\ell}$ (i.e., the linear interference term $p_{\mathrm{SL}}$) at the source~\cite{Falkowski:2019xoe}. The combined Fermi plus Gamow--Teller treatment has already been evaluated in the literature for the effect of scalar interactions on antineutrino production~\cite{Chaves:2021kxe}. A nonzero $p_{\mathrm{SS}}$ would activate the higher-order terms in second line of Eq.~\eqref{eq:PeeNSI}, leading to stronger constraints on the analyzed dataset. Also, as emphasized in  Ref.~\cite{Falkowski:2019kfn}, the QM-NSI formalism can break down beyond linear order in $\varepsilon_{e\ell}$, since consistency with the QFT description requires the matching conditions  $|d_{SL}|^{2}=d_{SS}$, is evidently not satisfied in the scalar channel considered here. We incorporate this structure directly in the full survival probability of Eq.~\eqref{eq:PeeNSI} used in our numerical analysis.

\section{Conclusion}
\label{sec:conclusion}

We investigated the sensitivity of JUNO to CC-SNSI within the QFT framework of neutrino oscillations. We derived an analytical expression for the electron antineutrino survival probability up to $\mathcal{O}(\varepsilon_{e\ell}^4)$, retaining scalar NSI contributions at detection, and showed that the resulting spectral distortions are degenerate with the solar oscillation parameters $\sin^2\theta_{12}$ and $\Delta m^2_{21}$, giving rise to a LMAs. We showed that this degeneracy is broken at $3\sigma$ with one year of data analyzing the solar sector. The quadratic term dominates once $|\varepsilon_{e\ell}|$ approaches order unity, as constrained by the 59.1- and 207.2-day JUNO datasets, so the full $\mathcal{O}(\varepsilon_{e\ell}^2)$ expression, not the linear truncation, is required to capture the true shape of the allowed regions. This is precisely where the QFT-based formalism goes beyond the linear approximation and beyond the reach of the QM-NSI formalism, which Ref.~\cite{Falkowski:2019kfn} showed to break down beyond linear order.

Using the first 59.1-day JUNO dataset, we obtained the first individual, rather than flavor-summed, neutrino-oscillation constraints on $\varepsilon_{e\mu}$ and $\varepsilon_{e\tau}$ from reactor antineutrinos, $|\varepsilon_{e\mu}|\le1.41$ and $|\varepsilon_{e\tau}|\le1.08$ at $1\sigma$, already competitive with existing reactor bounds despite the short exposure. The preliminary 207.2-day dataset tightens these limits to $|\varepsilon_{e\mu}|\le0.78$ and $|\varepsilon_{e\tau}|\le0.90$, confirming that the quadratic term, not the linear interference, closes the bulk of the allowed region. Projecting to ten years of JUNO operation, the sensitivity improves by more than an order of magnitude, $|\varepsilon_{e\mu}|\le5.1\times10^{-2}$ and $|\varepsilon_{e\tau}|\le3.6\times10^{-2}$, surpassing all existing determinations of these coefficients from reactor neutrino oscillations.

\begin{acknowledgments}
We thank the collaborators who contributed comments and discussions during the development of this manuscript. G.A.N. acknowledges support from the S\~ao Paulo Research Foundation (FAPESP) through Ph.D. scholarship No.~2024/01465-1. P.C. acknowledges support from the S\~ao Paulo Research Foundation (FAPESP) through Visitant professor fellowship No.~2024/20462-3. A.C.  acknowledges support from the National Council for Scientific and Technological Development – CNPq, project 446121/2024-0. PP is supported by Coordenação de Aperfeiçoamento de Pessoal de Nível Superior - Brasil (CAPES) - Finance Code 001, with grant PIPD-CAPES, number 33003017. O.L.G.P. acknowledges 
support from CNPQ through  306405/2022-9. 
\end{acknowledgments}

\appendix

\section{General CC-NSI and Matter Effects}
\label{app:cc-nsi}

In the QFT description of neutrino oscillations~\cite{Falkowski:2019xoe},
production, propagation, and detection are treated coherently at the amplitude level, so that charged-current NSI enter the rate through modifications of both the source and the detector vertices. The neutrino event rate normalized to the SM expectation reads
\begin{equation}
\frac{R_{ee}}{\phi_e^{\rm SM}\sigma_e^{\rm SM}} = \sum_{k,l} e^{-2i\Delta_{kl}}\, V_{e}^{kl}(p_X)\,\bigl[V_{e}^{kl}(d_X)\bigr]^{*},
\label{eq:rate_master}
\end{equation}
where $\Delta_{kl}\equiv\Delta m^2_{kl}L/4E_\nu$ as in Eq.~(\ref{eq:Pab_QFT}), the indices $k,l=1,2,3$ run
over mass eigenstates, and $X$ labels the considered CC-NSI Lorentz structure. The production factors are
\begin{align}
V_e^{kl}(p_X) &= U^{*}_{ek}U_{el}
+ p_{XX}\,(\varepsilon^X U)^{*}_{ek}\,(\varepsilon^X U)_{el} \notag\\
&+p_{XL}\left[(\varepsilon^X U)^{*}_{ek}\,U_{el} + U^{*}_{ek}\,(\varepsilon^X U)_{el}\right] , \label{eq:Vprod}
\end{align}
and analogously for $V_e^{kl}(d_X)$, replacing $p\to d$. Here $p_{XL,XX}$, are the NSI-to-SM amplitude ratios defined in the main text. Expanding Eq.~\eqref{eq:rate_master} with the oscillation amplitude $S_{\alpha\beta}$ defined in Sec.~\ref{sec:formalism} and the identity
\begin{equation}
S_{\alpha\gamma}^{*}\, S_{\beta\delta}=\sum_{k,l} e^{-i\Delta_{kl}}\, U_{\alpha k}^{*} U_{\beta l}\, U_{\gamma k} U_{\delta l}^{*},
\label{eq:Sidentity}
\end{equation}
collapses each cross term onto a product of oscillation amplitudes.
For a single non-vanishing parameter $\varepsilon_{e\ell}\neq 0$
($\ell=\mu,\tau$), the flavour sums in
Eqs.~\eqref{eq:rate_master} reduce the $d_{\rm XL}$ interference
term to the manifestly real form
\begin{equation}
\varepsilon^X_{e\ell}S_{ee}^{*}S_{e\ell}+\varepsilon^{X*}_{e\ell}S_{ee}S_{e\ell}^{*}
=2\,\Re\!\left(\varepsilon^{X}_{e\ell}S_{ee}^{*}S_{e\ell}\right),
\label{eq:interference_real}
\end{equation}
and likewise for the higher-order terms. When the mixed production
coefficient $p_{\rm XL}\neq 0$, as is relevant for tensor interactions, the
production factor in Eq.~\eqref{eq:Vprod} acquires the additional term
$p_{\rm XL}[(\varepsilon^X U)^{*}_{ek}U_{el}+U^{*}_{ek}(\varepsilon^X U)_{el}]$ compared with the CC-SNSI scrutinized in this work.

Beyond generalizing Eq.~\eqref{eq:PeeNSI} to other possible interactions, matter effects are included by the standard replacement of vacuum oscillation parameters by their effective, matter-modified counterparts, denoted with a tilde: $\theta_{12}\to\widetilde\theta_{12}$,
$\theta_{13}\to\widetilde\theta_{13}$, $\Delta m^2_{21}\to \Delta
\widetilde{m}^2_{21}$. Analytical expressions can be obtained following the perturbative approach of
Ref.~\cite{Khan:2019doq}, so that $U\to\widetilde U$ and
$S_{\alpha\beta}\to\widetilde S_{\alpha\beta}$. In this work, matter effects are added numerically, such that the complete transition rate reads

\begin{widetext}
\begin{align}
\frac{R_{ee}}{\phi_e^{\rm SM}\sigma_e^{\rm SM}} ={}& \widetilde P_{ee} + 2(d_{\rm XL}+p_{\rm XL})\, \Re\!\left(\varepsilon^{X}_{e\ell}\,\widetilde S_{ee}^{*} \widetilde S_{e\ell}\right) + \bigl(d_{\rm XX}+p_{\rm XX}+p_{\rm XL}d_{\rm XL}\bigr)\, |\varepsilon^{X}_{e\ell}|^2\,\widetilde P_{\ell e} \nonumber\\
&+ 2(p_{\rm XX}d_{\rm XL}+p_{\rm XL}d_{\rm XX})\,|\varepsilon^{X}_{e\ell}|^2\, \Re\!\left(\varepsilon^{X}_{e\ell}\,\widetilde S_{\ell e}^{*} \widetilde S_{\ell\ell}\right) + p_{\rm XX}d_{\rm XX}\,|\varepsilon^{X}_{e\ell}|^4\,\widetilde P_{\ell\ell}.
\label{eq:Ree_general}
\end{align}
\end{widetext}
Equation~\eqref{eq:Ree_general} is the matter-corrected CC-NSI rate used in
our  analysis, and holds for both scalar ($X=S$) and tensor
($X=T$) interactions, the two cases differing only through the values of their production and detection coefficients. For a purely
scalar interaction the Gamow--Teller restriction at production sets
$p_{\rm TL}$ and $p_{TT}$ to non-zero values, consequently the Eq.~\eqref{eq:Ree_general} does not reduces to the
simply $\mathcal{O}(\varepsilon^{2}_{e\ell})$ when the tensorial interaction is taken into account, but the full expression should be employed.

\addcontentsline{toc}{section}{{References}}
\bibliographystyle{peBibStyle}
\bibliography{biblio.bib}

\end{document}